\documentstyle[aps,prl,multicol,fancyheadings,epsfig,amsbsy,amssymb,amstex]{revtex}

\def\bbbc{{\mathchoice {\setbox0=\hbox{$\displaystyle\rm C$}\hbox{\hbox
to0pt{\kern0.4\wd0\vrule height0.9\ht0\hss}\box0}}
{\setbox0=\hbox{$\textstyle\rm C$}\hbox{\hbox
to0pt{\kern0.4\wd0\vrule height0.9\ht0\hss}\box0}}
{\setbox0=\hbox{$\scriptstyle\rm C$}\hbox{\hbox
to0pt{\kern0.4\wd0\vrule height0.9\ht0\hss}\box0}}
{\setbox0=\hbox{$\scriptscriptstyle\rm C$}\hbox{\hbox
to0pt{\kern0.4\wd0\vrule height0.9\ht0\hss}\box0}}}}
\newcommand{\beq}{\begin{eqnarray}} 
\newcommand{\eeq}{\end{eqnarray}} 

\begin{document}
\title{Competing Quantum Orderings in Cuprate Superconductors: A Minimal
Model}
\author{Ivar Martin, Gerardo Ortiz, A. V. Balatsky, and A. R. Bishop}
\address{Theoretical Division, Los Alamos National Laboratory, Los
Alamos, NM 87545}

\date{\today }

\maketitle

\begin{abstract}

We present a minimal model for cuprate superconductors.  At the unrestricted mean-field level, the model produces homogeneous superconductivity at large doping, striped superconductivity in the underdoped regime and various antiferromagnetic phases at low doping and for high temperatures.  On the underdoped side, the superconductor is intrinsically inhomogeneous and  global phase coherence is achieved through Josephson-like coupling of the superconducting stripes.  The model is applied to calculate experimentally measurable ARPES spectra.
\end{abstract}
\pacs{Pacs Numbers: XXXXXXXXX}

\vspace*{-0.4cm}
\begin{multicols}{2}

\columnseprule 0pt

\narrowtext
\section{Introduction}
 
We are witnessing an increase of experimental evidence indicating that
(charge and magnetic) incommensuration characterize the
low-energy physics of underdoped cuprate superconductors, both above and
below the critical superconducting temperature $T_c$ \cite{mason}.
This poses a challenging problem to theorists since these compounds
appear to be at the verge of a multitude of
different quantum ordered states that can be tuned by varying physical
parameters of the system. Theorists like to use the word ``quantum
criticality'' to refer to this phenomenon. The truth is, however, that
so far there is no rigorous theoretical framework that can explain
unambiguously this variety of complex phenomena (superconductivity,
magnetism, incommensuration, etc.), characterized by intrinsic
nonlinearities producing large-scale sensitivities to small
perturbations. 

In this paper we present a minimal model of high-$T_c$ superconductors
that clearly displays a variety of commensurate and incommensurate
competing thermodynamic phases. The advantage of our approach is that
it is simple, not subject to vague argumentation, and it allows one to
rigorously and exhaustively explore a variety of physical observables. 

Crucial to the experimental findings have been neutron scattering
techniques which probe the spin dynamics of the high-$T_c$ compounds
and suggest that different families of cuprate superconductors
share inhomogeneously spin and charge textured phases as their quantum
states \cite{mason}. It is well-known that the stoichiometric (half-filled) compounds are  antiferromagnetic (AF) Mott insulators as a result of strong electron
interactions and it is upon charge doping that they display
incommensuration.  Indeed, in a recent paper \cite{ours1} we have
presented a unified theory for the commensurate resonance peak and low-energy
incommensurate response observed in neutron scattering experiments. We
ascribe both features to be purely magnetic in origin: They represent
universal features signaling the existence of an incommensurate spin
state both below and above the superconducting transition temperature.
Our interpretation indicates that superconductivity {\em is not} the reason for the resonance peak, and that the incommensurate quantum state provides a reference state for the underdoped cuprates.

In previous work \cite{ours2,eARPES} we introduced two classes of microscopically {\it inhomogeneous} models which captured the magnetic and pairing properties of underdoped cuprates. Starting from a generalized $t$-$J$ model Hamiltonian in which appropriate terms mimic stripes, we found that inhomogeneous interactions that locally break magnetic $SU(2)$ symmetry can induce substantial pair binding of holes in the thermodynamic limit.  We showed that these models qualitatively reproduce the ARPES and neutron scattering data seen experimentally.  Moreover, based on the phenomenology of our
microscopic model we developed a mean-field (``Josephson spaghetti'')
model which provides a scenario for the macroscopic superconducting
state.  From our model Hamiltonian of random stripe separation $r$ and associated inter- and intra-stripe random Josephson coupling $J(r) \sim 1/r$ we obtained the experimentally observed relation $T_c(x) \simeq
\langle J(r) \rangle \propto [\langle r \rangle]^{-1} = \delta(x)$, where 
$x$ represents charge doping and $\delta$ is the inverse of a characteristic 
length scale associated with the incommensuration.

Our previous numerical simulations have helped to elucidate a certain fraction
of the underdoped cuprate puzzle.  Here we assume a different
strategy complementary to the previous approach: We propose a
minimal {\em homogeneous} model based on the one-band repulsive Hubbard Hamiltonian on a square lattice. The attractive particle-particle singlet channel is included through the nearest neighbor attraction $V$ \cite{aligia}, which produces predominantly $d$-wave pairing close to half-filling. We
solve this model at the mean-field level allowing all physical
quantities to vary from one lattice site to another.  In this way,
quantum morphologies characterized by a certain correlation length $\xi$
will appear when the minimum length of our supercell is larger than
$\xi$.

The basic question we address in this paper is whether
antiferromagnetic striped ordering \cite{Note1} and $d$-wave
superconductivity can coexist in a certain parameter range of our
model. In our opinion, this question is crucial for the understanding 
of the superconducting state in the underdoped materials. 

In the next Section we introduce the model and briefly describe the way
we solve it. We then summarize the resulting competing quantum states in
a phase diagram and discuss the spectral density. At the end of the paper, we review our main findings.

Some of the results presented in this paper have been introduced in 
Ref.~\cite{ours3}; the analysis of the spectral density is
described here for the first time.  

\section{Minimal Model}

We consider here a minimal model that brings together stripes and superconductivity.  The model is the two-dimensional one-band Hubbard  Hamiltonian with an on-site repulsion $U$ \cite{schulz}. Pairing correlations are introduced by including the nearest neighbor attraction $V$ \cite{micnas}. The effective minimal Hamiltonian is thus
\beq\label{eq:Httu}
H_{t-t^\prime-U} &=& - \sum_{i,j, \sigma}{ t_{ij}c^\dagger_{i\sigma}
c^{\;}_{j\sigma}} +  U \sum_{i}{n_{i\uparrow} n_{i\downarrow}}\\
\label{eq:H}
H &=& H_{t-t^\prime-U} + V\sum_{\langle ij \rangle}{n_i n_j},
\eeq
where the operator $c^{\dagger}_{i\sigma}$  ($c^{\;}_{j\sigma}$) creates  (annihilates) an electron with spin $\sigma$ on the lattice site $i$, and $n_i = c^{\dagger}_{i\uparrow}c^{\;}_{i\uparrow} + c^{\dagger}_{i\downarrow}c^{\;}_{i\downarrow}$ represents the electron density on site $i$. The hopping $t_{ij}$ equals $t$ for nearest neighbors and $t^\prime$ for for the second-nearest neighbor sites $i$ and $j$.  For our computations, we use the unrestricted mean-field approximation to this Hamiltonian,
\beq\label{eq:H_MF}
H_{MF} = -\sum_{\langle ij \rangle \sigma}{ t_{ij} c^\dagger_{i\sigma}
c^{\;}_{j\sigma}} + U \sum_{i}{n_{i\uparrow} \langle n_{i\downarrow}
\rangle + \langle n_{i\uparrow}\rangle n_{i\downarrow}} \nonumber\\ +
\sum_{\langle ij \rangle}{c^{\;}_{i\downarrow} c^{\;}_{j\uparrow}
\Delta_{ij}^* + {\rm H.c.}} \ ,
\eeq
where $\Delta_{ij} = V \langle c_{i\downarrow} c_{j\uparrow}\rangle$ is the MF superconducting order parameter. The effect of $V$ in our model is limited to the generation of superconducting correlations. We do not explicitly address the important issue of the  microscopic origin of the attraction $V$ \cite{aligia}. 
\begin{figure}[htbp]
  \begin{center}
   \includegraphics[width = 3.0 in]{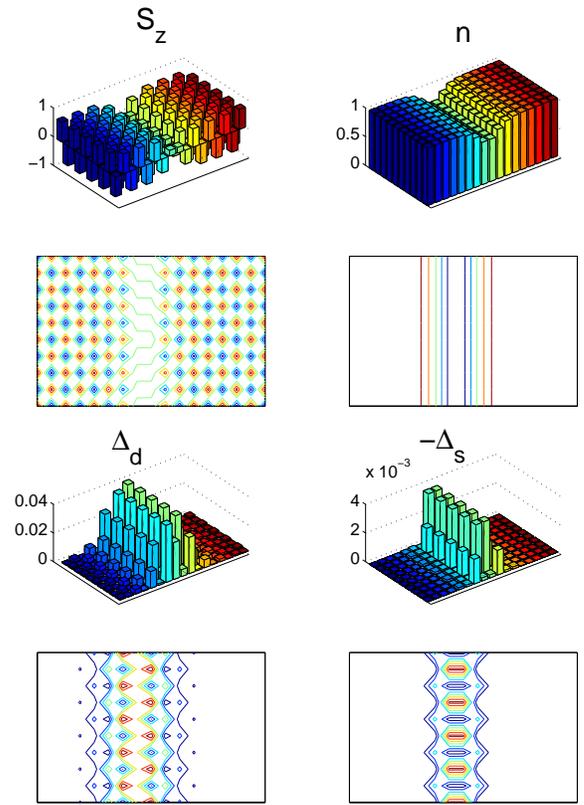}
\vspace{0.5cm}
\caption{Typical example of density and superconducting order parameter profiles in a stripe state (here, period 17).  The top two bar charts represent the site-dependent spin and charge densities, respectively. The contour plots indicate the sites with low (blue) and high (red) values of the corresponding densities. The bottom four plots show the values of the superconducting order parameters, defined as $\Delta_i^{d(s^*)} = (\Delta_{i, right} + \Delta_{i, left}  \mp\Delta_{i, up} \mp \Delta_{i, down})/4$  for $d$-wave (extended s-wave)  order parameter on site $i$ ($U = 4t$, $V = -0.9t$, $t^\prime = 0$). Different choices of parameters lead to qualitatively similar patterns, with stronger $U$ leading to a stronger AF order and more attractive $V$ causing the superconducting stripes to become wider and larger in amplitude. The doping level is $5.9\%$.}
\label{fig:10x17}
\end{center}
\end{figure}

A typical zero-temperature MF inhomogeneous solution is shown in Fig. \ref{fig:10x17}. In the lowest energy configuration, the spin density develops a soliton-like AF anti-phase domain boundary --- a stripe --- at which the AF order parameter changes sign. At the domain boundary, the electronic charge density is depleted. The width of the domain wall, $\xi_{DW}$, decreases with increasing on-site repulsion $U$. However, for values of $U$ that are not much larger than the hopping $t$, the charge per unit length of the optimal (the
lowest energy) stripe remains the same and is close to unity near half-filling for $t^\prime = 0$.  The bond-centered stripes are favored relative to the site-centered ones, although the energy difference in our case is small due to the smooth charge distribution.  

Stripe formation is the
result of the competition between antiferromagnetism (which can lead to
charge confinement) and delocalization (driven by kinetic energy). 
Non-linear feedback is responsible for these complex patterns.  A
half-filled (one electron per site) antiferromagnet is the state with
the lowest energy per electron. Upon small doping, the energy can be
optimized by segregating the excess electrons or holes into domain
walls or charged clusters, and keeping the bulk antiferromagnetism
unperturbed. The anti-phase domain walls are favored over the simple
charged stripes as they allow the charge carriers to optimize their
transverse fluctuations in the direction across stripes, thereby
lowering their kinetic energy. The linear filling of the emergent
stripes vary depending upon the specific nature of the band
structure.
\begin{figure}[htbp]
  \begin{center}
   \includegraphics[width = 3.0 in]{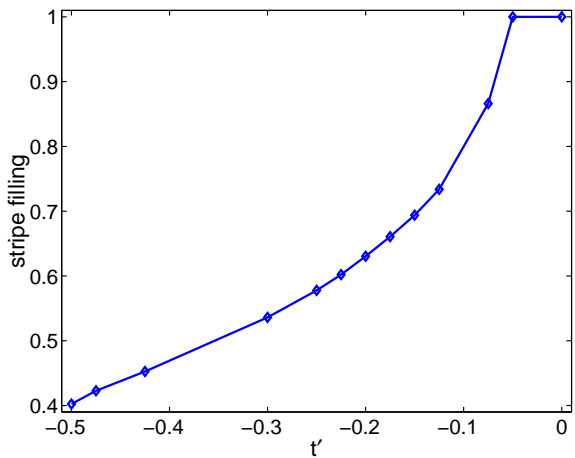}
\vspace{0.3cm}
\caption{Linear filling of an isolated vertical stripe as a function of next-nearest neighbor hopping, $t^\prime$.  Here $U = 4 t$ and $V = 0$.}
\label{fig:tprime}
\end{center}
\end{figure}

For different band structures the exact relation between the doping $x$ and inter-stripe distance, $L(x)$, may change; however, any model whose ground state is AF at zero doping, can be expected to have AF stripes for a finite doping, with incommensuration proportional to the doping, $1/L(x) \propto x$, near half-filling.  For example, negative next-nearest neighbor hopping $t^\prime$ (relevant in the hole-doped cuprates\cite{ws}), modifies the stripe filling without compromising the stripe phase stability relative to commensurate AF at the MF level\cite{machida2}.  The stripe filling is a monotonically decreasing function of the magnitude of $t^\prime$, with the filling 1/2 occurring when 
$t^\prime = -0.35 t$ (Figure \ref{fig:tprime}).  
While filling-one stripes correspond to a correlated insulator, the fractional filling stripes in the $t$-$t^\prime$-$U$ model (Eq.~(\ref{eq:Httu})) are metallic, which can be understood in terms of the partial occupancy of the mid-gap band formed due to the stripes\cite{machida2}.  In the case of insulating stripes ($t^\prime = 0$) a threshold value of attraction $V$ should be exceeded to generate superconductivity and hence to overcome the insulating gap\cite{zaanen_SC}; however, in the case of metallic stripes one would expect that any attraction would yield superconductivity through the Cooper instability\cite{BCS}.  Indeed, this is what we find \cite{ours3}.
On the contrary, the diagonal stripes which can also be the ground states of the Hubbard model, particularly at low dopings, always have a filling of one electron per Cu site, and hence are insulators.  This makes diagonal stripes antagonistic to superconductivity \cite{machida2}, and also agrees with the experimental observations \cite{wakimoto}.

From Figure \ref{fig:10x17} it is clear that the superconducting order parameter $\Delta_{ij}^{d(s^*)}$ is maximized on the stripes and is not smooth (even within the stripe) due to the presence of the AF background. In addition to the dominant $d$-wave component, there is a small extended $s$-wave ($s^*$) component generated on the stripe, which can be interpreted as a distortion of the $d$-wave at the level of about $10\%$. This happens because the symmetry of the lattice has been spontaneously broken by the stripes.  For dopings less than about $10\%$ (corresponding to $L(x) > 10$ lattice sites) the stripes have negligible overlap . In this regime, the amplitude of the superconducting order parameter on the stripes no longer depends upon the stripe-stripe separation. For higher doping levels, an overlap between the superconducting order parameters on adjacent stripes is established, and for even higher doping the stripes ``melt'' and superconductivity becomes homogeneous, of a classical BCS type \cite{BCS}.

\subsection{\underline{Phase diagram}}
\begin{figure}[htbp]
  \begin{center}
   \includegraphics[width = 3.0 in]{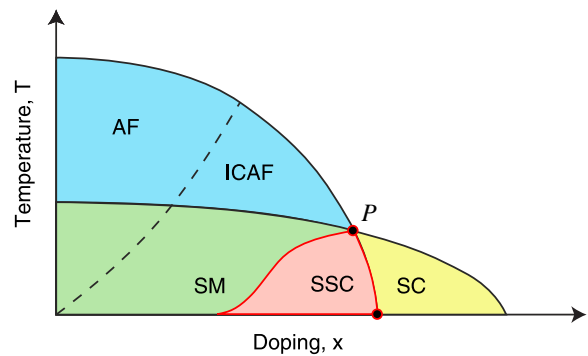}
\vspace{0.3cm}
\caption{Schematic phase diagram obtained by superimposing the antiferromagnetic (AF) / striped (ICAF) and the $d$-wave superconducting (SC) phase diagrams. In the intersection region we distinguish the subregions of  Josephson-coupled striped superconductor (SSC), and non-superconducting ``strange metal'' (SM),  which is neither a superconductor, nor a simple insulator.  The upper boundary  of the AF/ICAF corresponds to the weak pseudogap crossover, and the line  between the pure AF/ICAF and the SM marks the strong pseudogap  crossover.  A detailed finite-temperature study is required to precisely locate the left boundary of the SSC region, and hence to determine the order of the critical point $P$.}
\label{fig:pd}
\end{center}
\end{figure}

From our zero-temperature analysis of the coexistence of AF stripes (ICAF) and superconductivity \cite{ours3}, a simple qualitative thermodynamic phase diagram emerges. In the conjectured phase diagram, we utilize the finite-temperature AF/ICAF phase diagram of the Hubbard model together with the superconducting (SC) phase diagram of the $t$-$V$ model. The SC phase diagram is obtained in the homogeneous MF \cite{micnas}, while the AF/ICAF phase boundary is constructed under the assumption of the second order phase transition between the homogeneous and inhomogeneous states \cite{schulz}.
For a suitable choice of parameters, for instance $U = 2t$ and $V = - t$, the 
SC and the AF/ICAF regions in the phase diagram intersect, as shown in Fig. \ref{fig:pd}.  The energy scale associated with the AF/ICAF region of the phase diagram is much larger than that of the SC part. Thus, one expects that only the SC phase boundary is modified when it passes through the AF/ICAF region. The central result of our work is that the superconductivity {\em does not} disappear in the region of the AF stripes, but rather becomes striped, with anisotropic superfluid stiffness.

Based on familiar Josephson coupling physics, in the region of coexistence of superconductivity and stripes, we can expect a part that is a globally coherent striped superconductor (SSC). The rest of the intersection region is covered by an exotic phase  which, if it were perfectly orientationally ordered, would be a superconductor in one direction and a strongly-correlated insulator in the other. In reality, due to the meandering of the stripes and their break-up into finite segments \cite{fradkin}, the state is likely to be highly inhomogeneous and neither an insulator, nor a superconductor, but also not a simple metal. In agreement with the experimental attribution, we refer to this region as a ``strange metal'' (SM). The line separating the SM from the rest of the AF/ICAF region, in the context of the experiments, can be associated with the crossover to the strong pseudogap regime, and corresponds to the opening of the local superconducting gap. The high-temperature boundary separating AF/ICAF phases from the homogeneous state, marks the onset of the weak pseudogap.  For small dopings, there is also a possibility of a transition from the vertical to diagonal stripes \cite{machida2}.

\subsection{\underline{Spectral Density}}
In this Section we demonstrate how our model can be applied to compute the energy spectrum of the system, which can then be compared with the experimental data \cite{zxshen}.  At any temperature, the MF solution yields a self-consistent spectrum $\{E_n\}$ of Bogoliubov quasiparticles, which diagonalize the MF Hamiltonian in Eq. (\ref{eq:H_MF}).  Knowing how the electron operators are related to the Bogoliubov quasiparticles, one can compute the electronic spectral densities for positive (particle) and negative (hole) biases.  
Experimentally, angle resolved photoemission (ARPES) measures the electronic spectral density integrated in the window of $\pm \Delta \varepsilon$ around the Fermi energy.  

\begin{figure}[htbp]
  \begin{center}
   \includegraphics[width = 3.0 in]{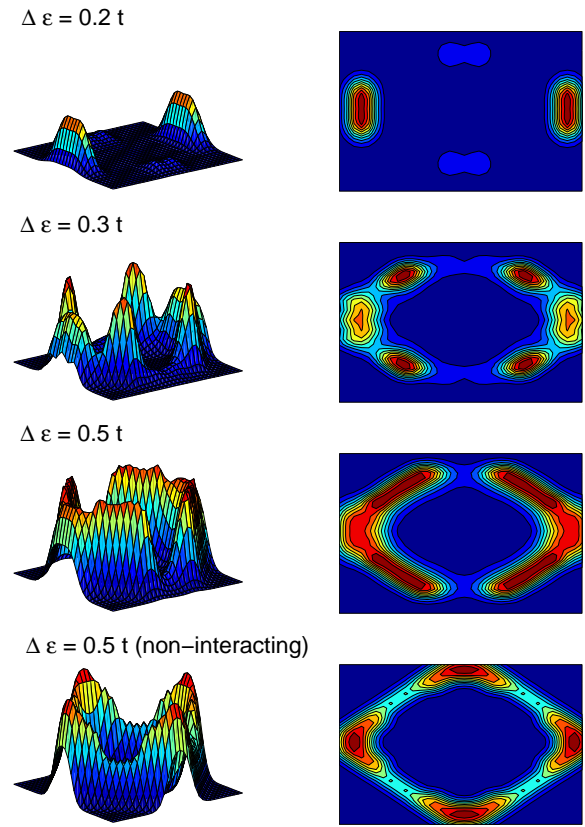}
\vspace{0.5cm}
\caption{Computed ARPES spectra for various energy integration windows, $\Delta \varepsilon$.  The Brillouin zone is defined as $(0, 2\pi)\times(0,2\pi)$, with the horizontal direction being along the stripes and the vertical direction perpendicular to the stripes.  Notice that for small  $\Delta \varepsilon$ the spectral weight is concentrated around $(0,\pi)$, with the Fermi surface being gradually reconstructed with the increasing window of integration.  The parameters are $U = 4 t$, $V = -t$, $t^\prime = -0.2 t$, and doping 8.3\%.  The system size is $16\times 16$, with two collinear stripes.  For comparison we also show the case of free electrons with the same non-interacting band structure and doping.}
\label{fig:arpes}
\end{center}
\end{figure}

In Figure \ref{fig:arpes} we show the computed ARPES spectra for various energy integration windows.  The symmetry of the spectrum is spontaneously broken due to the presence of stripes (the stripes run along the horizontal $k$ direction).  For small integration window near the Fermi surface, the spectral weight is concentrated around the $(0, \pi)$ point.  The reason is that the stripes gap out the flat parts (``diagonals'') of the Fermi surface, while keeping the quasiparticles around $(0, \pi)$ gapless if $t^\prime < 0$ or weakly gapped if $t^\prime = 0$.  Since these are the quasiparticles that are primarily responsible for the formation of $d_{x^2-y^2}$ superconductivity, it is this particular structure of the ``stripe gap'' that allows for the peaceful coexistence of stripes and superconductivity in our model.
Indeed, in the absence of superconductivity ($V = 0$), the spectral patterns remain essentially unchanged, except for the enhanced weight around $(0, \pi)$.

For larger energy integration windows, the Fermi surface gradually ``reconstructs,''  with the energy states around the diagonal reappearing when $\Delta \varepsilon$ exceeds the ``stripe gap.''

Notice that due to twinning and the expected presence of stripe domains in the real experimental systems the computed spectra have to be symmetrized.  Similar results for the striped spectra (but without superconductivity) have been obtained previously \cite{kARPES,machida2,eARPES}.  Similarly, one can calculate temperature-dependent specific heat, entropy, spin susceptibility, among other experimentally measurable quantities, as we will report elsewhere.

\section{Conclusions}
In summary, we have presented a minimal model supporting the
coexistence of incommensurate antiferromagnetism (``stripe'' order)
and global anisotropic superconductivity.  Contrary to the common belief, these two order parameters can coexist and our calculation is a faithful realization of such a physical situation.  
At the same time, the stripe order provides a natural competing order parameter limiting the increase of the superconducting transition temperature on the underdoped side of cuprates.

Our model displays a variety of other competing
homogeneous and inhomogeneous thermodynamic phases.  Based on the model, we constructed a phase diagram that captures many features of the superconducting cuprates.  Finally, we computed the photoemission spectra which give a clear interpretation of experimental data.

This work was supported by the U.S. DOE.

\end{multicols}
\end{document}